\documentclass[onecolumn,showpacs,preprintnumbers]{revtex4}
\usepackage{amsfonts}
\usepackage{amsmath}
\usepackage{graphicx}

\input{tcilatex}
\begin{document}

\date{\today }
\title{Stagflation --- Bose-Einstein condensation in the early universe}
\author{Takeshi Fukuyama \thanks{%
fukuyama@se.ritsumei.ac.jp}}
\affiliation{Department of Physics and R-GIRO, Ritsumeikan University Kusatsu, Shiga,
525-8577 Japan }
\author{Masahiro Morikawa \thanks{%
hiro@phys.ocha.ac.jp}}
\affiliation{Department of Physics, Ochanomizu University, 2-1-1 Ohtuka, Bunkyo, Tokyo
112-8610, Japan}

\begin{abstract}
Our universe experienced the accelerated expansion at least twice; an
extreme inflationary acceleration in the early universe and the recent mild
acceleration. By introducing the Bose-Einstein condensation (BEC) phase of a
boson field, we have been developing a unified model of dark energy (DE) and
dark matter (DM) for the later mild acceleration. In this scenario, two
phases of BEC (=DE) and normal gas (=DM) transform with each other through
BEC phase transition. This unified model has successfully explained the mild
acceleration as an attractor. We extend this BEC cosmology to the early
universe without introducing new ingredients. In this scenario, the
inflation is naturally initiated by the condensation of the bosons in the
huge vacuum energy. This inflation and even the cosmic expansion eventually
terminates exactly at zero energy density. We call this stage as
stagflation. At this stagflation era, particle production and the decay of
BEC take place. The former makes the universe turn into the standard hot big
bang stage and the latter makes the cosmological constant vanishingly small
after the inflation. Furthermore, we calculate the density fluctuations
produced in this model, which turns out to be in the range allowed by the
present observational data. We also show that the stagflation is quite
robust and easily appears when one allows negative region of the potential.
Further, we comment on the possibility that BEC generation/decay series
might have continued all the time in the cosmic history from the inflation
to present.
\end{abstract}

\maketitle



\section{Introduction}

Accelerated expansion of the universe seems not to be exceptional.
Theoretical and observational studies have revealed that the universe
experienced the era of accelerated cosmic expansion at least twice; an
extreme inflationary acceleration in the early universe (EA)\cite%
{Tsujikawainflation} and a late mild acceleration (LA)\cite{Fukuyama96} \cite%
{TsujikawaDE}. We would like to figure out the basic physics behind these
accelerations and reveal the inevitability, if any, of these accelerations.
In the context of the Einstein equation, we need special matter which is
endowed with strong negative pressure in order to guarantee the accelerated
cosmic expansion.

However, it is often very difficult to find such matter in baryons, and
therefore new fields such as the \textsl{classical scalar field} has been
hypothesized though we have no fundamental understanding of it. Probably
reflecting this uncertainty, this field has been named differently in
various contexts in the literature such as Higgs field, inflaton in EA, and
K-essence, tachyon, quintessence, Chaplygin gas, phantom field, ghosts, etc.
in LA.

We do not want to be involved in these particular contexts but instead, we
would like to reconsider how the degrees of freedom of such scalar field can
arise within a firm physics even if we still cannot fully identify this
field. To this aim, we reconfirm the fact that the basic physics which
describes the universe from the fundamental level is no doubt the quantum
field theory. Then how the classical dynamical degrees of freedom of scalar
field can arise within this theory? The most natural mechanism we can
consider would be the \textsl{Bose-Einstein condensation} (BEC) of some
boson field\cite{Pethick08}. This condensation process is a phase transition
and a typical mechanism that a classical degrees of freedom emerges as an
order parameter within the quantum field theory \cite{Morikawa07}.

Along this line of thought, we have developed a cosmological scenario based
on BEC in order to explain LA in the contemporary universe\cite{Nishiyama04} 
\cite{Fukuyama06} \cite{Fukuyama07} . In this scenario, the dark energy (DE)
is identified with the BEC of the bosons and the dark matter (DM) with the
excited gas of the bosons. This cosmological situation is analogous to the
liquid $^{4}He$ system in the laboratory in the sense that the two
distinguished components, \textsl{super and normal, coexist and interchange
with each other}. Our unified model of DE and DM has been successful to
describe the contemporary universe so far.

In this paper, we extend this BEC cosmology to the early universe in order
to examine how our BEC cosmology reveal another acceleration EA. Since we
treat LA and EA on an equal footing, we should not introduce any new
ingredients into the previously developed BEC cosmology except for the
cosmological background. Thus the goal of this paper is to examine to what
extent this BEC cosmology can be verified when applied to the early
universe. We are especially interested in how the mechanisms of LA and EA
are different with each other in BEC cosmology. We remember that LA,
described in BEC cosmology, has been an attractor \cite{Fukuyama06} \cite%
{Fukuyama07} which guarantees its autonomous realization. Can we also expect
EA be an autonomous mechanism in BEC cosmology?

More precisely, when we analyze EA, we should clarify the \textsl{autonomous
initiation and termination processes of the inflationary era}, which
naturally turns into\textsl{\ the reheated }radiation dominated era.
Actually, the BEC makes the order parameter develop from zero to some finite
value, and therefore the inflationary dynamics with finite duration would
initiate autonomously if sufficient vacuum energy is provided by quantum
fluctuations or something. We would verify this mechanism first. That is, we
derive BEC production rate from the microscopic Lagrangian by using the
generalized effective action in the in-in formalism. As for the termination
of the inflationary era, it may seem to be impossible to control the ever
increasing order parameter since our BEC model assumes the negative quartic
interaction (attractive force) without bound. However it is often seen in
many physical phenomena that a violently unstable mode is taken over by an
another stable mode via complex nonlinear processes. We will show that this
is the case for BEC cosmology. Actually an extreme development of the order
parameter completely prevents the violent cosmic expansion. We call this
process as \textsl{stagflation, }where the order parameter ever inflates
while the cosmic expansion becomes stagnant. Thus the inflationary era
terminates autonomously within a finite time scale. Moreover, this extreme
development of the ever accelerating order parameter easily gives rise to
the particle production, which provides a natural process of the reheating.
Furthermore, the instability analysis reveals that the homogeneous mode
immediately decays into the inhomogeneous localized modes at this
stagflation era, when the cosmic expansion ceases.

This stagflation era also provides a key mechanism to solve the \textsl{%
cosmological constant problem}. The issue is to explain the dynamics of the
cosmological 'constant' $\Lambda $, which had once been a huge value $%
(10^{21-26}$ eV$)^{2}$ that makes EA, turns into a tiny but non-zero value
about $(10^{-31}$ eV$)^{2}$ that makes LA. Though this initial huge value
for $\Lambda $\ would be naturally provided from ubiquitous vacuum energy,
there should be a certain mechanism for the very fine tuning of $\Lambda $
to the vanishingly small but finite value at present. In the previous
analysis of BEC cosmology applied to LA, we assumed that the zero-point of
the potential energy is exactly zero, and shown that the vanishingly small
but finite value of the effective $\Lambda $ term appears as a fixed point,
in which the condensation speed and the potential force balance with each
other. In this paper, we propose a new mechanism which autonomously adjust
the effective $\Lambda $ term exactly zero. This would justify the above
assumption for LA. The essence of this $\Lambda $-extinction mechanism is
the fact that \textsl{the stagflation takes place exactly at the zero total
energy}. As explained above, the uniform mode of BEC becomes unstable at
this stagflation point and suddenly decays into localized objects. Thus
after the inflation, the universe evolves into the stagflation point, where $%
\Lambda $ becomes zero and then the universe becomes dominated by the
reheated radiation. We would like to quantitatively verify this mechanism in
this paper. It will turn out that the $\Lambda $-extinction is
asymptotically realized after many BEC decays.

The above instability of the uniform mode is the same as that appears in BEC
cosmology application in LA. Moreover, the condensation process of BEC
commonly triggers all the essential dynamics of the cosmic history in both
the cases of EA and LA. It is important to notice that these decay due to
instability and the condensation of BEC are not a coherent time change
described by any Lagrangian but are rather phase transitions and the
incoherent evolution with dissipation and fluctuations. This point is
further discussed in the following sections.

This paper is organized as follows. After applying the in-in formalism into
BEC formation, we review the essence of BEC cosmology applied in LA in
section 2. In section 3, we apply the BEC cosmology to the early universe in
the order of initiation, termination of inflation, reheating, cosmological
constant problem, and generation of density fluctuations. In the last
section 4, we summarize our study emphasizing the robust stagflation and
clarifying unsolved problems at present.

\section{Basics of the BEC cosmology}

\subsection{BEC in the Universe}

In this section, we first review our model of BEC cosmology from the
microscopic physics. Various observations and theoretical analysis have
revealed that the DE and DM dominate the matter contents of the universe and
DE/DM ratio is of order unity at present\cite{TsujikawaDE}. These facts may
indicate some deep relation between DE and DM. Therefore it seems natural to
construct a unified model of DE and DM based on a single physical principle.
The simplest thought will be to consider that both DE and DM are actually
the same dynamical matter but in different phases of existence. DE phase
should be almost uniformly distributed and is supposed to provide
accelerated cosmic expansion. On the other hand, DM phase should be
inhomogeneously distributed and is supposed to trace the structure of
galaxies and clusters.

According to the Einstein equation, the acceleration of the cosmic expansion
requires isotropic negative pressure comparable to the energy density $%
p<-\rho /3$\ for DE. This is because, in general relativity, pressure as
well as energy density works as a source of gravity. Such strong negative
pressure sounds strange if we consider fermionic fields which constitute
ordinary baryonic matter. On the other hand for bosons, familiar static
magnetic field $\overrightarrow{H}$, for example,\ shows such behavior $%
p=-\rho =-\overrightarrow{H}^{2}/\left( 8\pi \right) $ in one spatial
direction, reflecting that the magnetic force line has a positive tension.
This negative pressure would be the origin of the variety of bursts and
accelerations in solar physics such as flares and coronas. Unfortunately
this is a vector field and shows positive pressure in the other two spatial
directions. A natural possibility to obtain the isotropic negative pressure
is to consider the classical scalar field. This would be the standard line
of thought in the modern cosmology as explained in the introduction.

However, when we construct a unified model of DE and DM, we further have to
specify the origin of such scalar field in order to clarify the connection
between DE and DM. The problem is to find a phase which behaves as classical
scalar field in the matter which is originally described by quantum physics.
The most natural case would be the quantum condensation of bosons in low
temperature, \textsl{i.e.} the Bose-Einstein condensation. Macroscopic
number of boson particles share a single ground state in common and behaves
as a coherent wave. This condensed phase is well described by a classical
mean field. On the other hand the rest of the boson gas in excited states
should macroscopically behaves as classical gas. Therefore we naturally
arrive at the thought that the boson gas is composed of two phases,
condensed phase and gas phase, each of which can be regarded as DE and DM,
respectively. Thus the BEC cosmology arises. Since the basic structure of
the universe in this model is quantum physical and cosmological, we have a
unique mass scale $m=\left( H_{0}^{2}h^{3}/\left( Gc^{3}\right) \right)
^{1/4}=0.0092$eV\ obtained by equating the present energy density $\rho
=H_{0}^{2}/G$\ to the energy density associated with mass $m$\ and the
Compton wave length $l=h/\left( mc\right) $. This scale characterizes the
cosmic BEC.

The above two-phase structure is quite common in physics in the sense that
any special interactions among boson particles are not required for BEC and
it inevitably takes place provided the temperature is lower than the
critical one. A well known such structure would be the liquid $^{4}He$
system in the laboratory. The super phase and normal phase correspond,
respectively, to DE and DM; they coexist and interchange with each other.
Although the origin of the two-phase is quantum phase transition, the
condensed phase is a classical mean field and should not be confused with
the macroscopic quantum state, which is described by a wave function.

The BEC of boson mass $m$ takes place provided the de Broglie wave length $%
\lambda _{dB}\equiv \sqrt{2\pi \hbar ^{2}/(mkT)}$ exceeds the mean
separation $n^{1/3}$

\begin{equation}
kT<kT_{cr}\equiv \frac{2\pi \hbar ^{2}n^{2/3}}{\left( \zeta \left(
3/2\right) \right) ^{2/3}m}.  \label{criticalTforBEC}
\end{equation}%
This condition requires the boson gas to be 'cold', if we were to realize
BEC for reasonable boson mass appropriate for DM. Actually in the universe,
if the boson is the cold dark matter, irrespective of its interaction, its
temperature (or quasi-temperature) behaves as the same as above $T\propto
n^{3/2}$ in the cosmic expansion. Therefore BEC can always take place and
continue provided the boson temperature was less than the critical
temperature at some moment of cosmic evolution. The corresponding critical
boson mass for BEC can be arbitrary in general, but if we temporary assume
that the boson was in equilibrium with the rest of the matter in the past,
then the upper limit of the mass turns out to be about $19$eV \cite%
{Fukuyama07}.

In the universe the ratio of DE and DM cannot simply be given by the
standard expression for the ratio of condensation $1-\left( T/T_{cr}\right)
^{3/2}$ in equilibrium with the temperature $T.$ The universe is actually
almost adiabatic and therefore the transition toward the ground state is
forbidden in the first approximation. However a small non-adiabaticity
allows the very slow condensation. We need a firm formulation which
describes this condensation process based on microscopic physics. The most
appropriate, but not necessarily complete, method for this purpose would be
the generalized effective action formalism \cite{Morikawa86}, \cite%
{Morikawa95}, \cite{Stoof99}, on which we now briefly explain.

The partition function, which describes all the fluctuations, is given by 
\begin{equation}
\tilde{Z}[\tilde{J}]\equiv tr\left[ \tilde{T}\left( \exp \left[ i\int_{c}{%
\tilde{J}\tilde{\phi}}\right] \rho \right) \right] \equiv \exp \left[ i%
\tilde{W}[\tilde{J}]\right] ,
\end{equation}%
where the integration is taken over the entire space and the \textsl{doubled
time contour }which extends from $t=-\infty $ toward $t=+\infty $ and then
back to $t=-\infty $ again. The symbol $\tilde{T}$ rearranges the operators
on its right hand side in the order of this doubled time contour. All the
quantities with overtilde represents that they are defined on this contour. $%
\tilde{J}$ is the external source field, $\rho $ is the initial density
matrix, and the quantum field $\phi $ is in the Heisenberg representation.
This partition function can also be represented in the interaction picture as

\begin{eqnarray}
\tilde{Z}[\tilde{J}] &=&tr\left[ \tilde{T}\left( \exp \left[ i\int \tilde{J}%
\tilde{\phi}-i\int V\left( \phi \right) \right] \rho \right) \right] \\
&=&\exp \left[ \tilde{T}\int V\left[ \frac{\delta }{i\delta \tilde{J}}\right]
\right] \exp \left[ -\frac{i}{2}\iint \tilde{J}\left( x\right) \tilde{G}%
_{0}\left( x,y\right) \tilde{J}\left( y\right) \right] \times \\
&&tr\left[ :\exp \left[ i\int \tilde{J}\tilde{\phi}\right] :\rho \right] 
\notag
\end{eqnarray}%
where the Lagrangian is decomposed as the free part $L_{0}[\phi ]$ and the
interaction part $V[\phi ]$, $L[\phi ]=L_{0}[\phi ]-V[\phi ]$. The Wick
theorem is used for the last equality. It is useful to introduce a vector
and matrix representation for the functions on the generalized time contour;
a field $\tilde{X}(x)$ can be written as $X_{+}(x)+X_{-}(x)$, where the
quantity with suffix $+(-)$ has its support only on the forward
(respectively, backward) time contour. Then $\tilde{X}(x)$ can be
represented as a vector $\left( X_{+}(x),X_{-}(x)\right) $. For example $%
\tilde{J}\left( x\right) =\left( J_{+}(x),J_{-}(x)\right) $ and $\tilde{\phi}%
(x)=\left( \phi _{+}(x),\phi _{-}(x)\right) $. We also use the notation $%
X_{\Delta }(x)\equiv X_{+}(x)-X_{-}(x)$ and $X_{C}(x)\equiv \left(
X_{+}(x)+X_{-}(x)\right) /2$ in the following. Then the notation $\int {%
\tilde{J}\tilde{\phi}}$ means $\int {\tilde{J}\tilde{\phi}}=\int {%
dxJ_{+}(x)\phi _{+}(x)}-\int {dxJ_{-}(x)\phi _{-}(x)}$, where the
subtraction on the right hand side represents the fact that the direction is
opposite in the backward time contour. {This can be conveniently expressed
in the two-by-two matrix notation, }%
\begin{equation}
\int {\tilde{J}\tilde{\phi}}=\int {dx}({J_{+}(x),J_{-}(x)})\sigma _{3}({\phi
_{+}(x),\phi _{-}(x)})^{T},
\end{equation}%
where $\sigma _{3}=\left( 
\begin{tabular}{cc}
$1$ & $0$ \\ 
$0$ & $-1$%
\end{tabular}%
\right) $ and $T$ is the transpose.

The generalized two-point function $\tilde{G}_{0}(x,y)\equiv -i\left\langle 
\tilde{T}{\phi (x)\phi (y)}\right\rangle $ is also expressed in this matrix
notation as 
\begin{equation}
\tilde{G}_{0}(x,y)=\left( 
\begin{tabular}{cc}
$G_{F}\left( x,y\right) $ & $G_{+}\left( x,y\right) $ \\ 
$G_{-}\left( x,y\right) $ & $G_{\bar{F}}\left( x,y\right) $%
\end{tabular}%
\right) =\left( 
\begin{tabular}{cc}
$-i\left\langle {T\phi (x)\phi (y)}\right\rangle $ & $-i\left\langle {\phi
(y)\phi (x)}\right\rangle $ \\ 
$-i\left\langle {\phi (x)\phi (y)}\right\rangle $ & $-i\left\langle {\bar{T}%
\phi (x)\phi (y)}\right\rangle $%
\end{tabular}%
\right) .
\end{equation}%
The fact that the only three of them are independent is most clearly
observed in their momentum representations, 
\begin{equation}
\hat{G}_{F}(k)=\frac{-D(k)+iB(k)}{D(k)^{2}+A(k)^{2}},\hat{G}_{\bar{F}}(k)=%
\frac{-D(k)-iB(k)}{D(k)^{2}+A(k)^{2}},\hat{G}_{\pm }(k)=i\frac{A(k)\mp B(k)}{%
D(k)^{2}+A(k)^{2}}.
\end{equation}%
where $A\left( k\right) ,B\left( k\right) ,$ and $D\left( k\right) $ are
mutually independent functions.

If we were to define the classical field as $\tilde{\varphi}(x)\equiv \delta 
\tilde{W}/\delta \tilde{J}$ and define the Legendre transform $\tilde{\Gamma}%
[\tilde{\varphi}]\equiv W[\tilde{J}]-\int {\tilde{J}\tilde{\varphi}}$\textbf{%
\ }as the effective action as usual, then by definition, the equality holds $%
\delta \tilde{\Gamma}/\delta \tilde{\varphi}(x)=-\tilde{J}(x)$, which may be
regarded as the equation of motion for $\tilde{\varphi}(x)$ including all
the quantum fluctuations. However, reflecting the fact that the effective
action is not in general real $\tilde{\Gamma}=Re\Gamma +iIm\Gamma $, the
above method makes the field $\tilde{\varphi}(x)$ unphysically complex. This
ordinary procedure cannot describe the phase transition dynamics. Therefore
we go beyond the standard method and decompose the effective action. We
first notice that $Im\Gamma $ is even in the variable $\varphi _{\Delta
}(x)\equiv \varphi _{+}(x)-\varphi _{-}(x)$, 
\begin{equation}
Im\tilde{\Gamma}=\frac{1}{2}\int \int {\varphi _{\Delta }(x)B(x-y)\varphi
_{\Delta }(y)dxdy+...}
\end{equation}%
and the kernel $B(x-y)$ is positive definite\cite{Morikawa86}. Therefore,
introducing an auxiliary real field $\xi (x),$ we have the following
path-integral expression. 
\begin{equation}
e^{i\tilde{\Gamma}[\varphi ]}=\int {[d\xi ]}P[\xi ]\exp [iRe\Gamma +\int
i\xi \left( x\right) \varphi _{\Delta }\left( x\right) dx]
\end{equation}%
where the weight function is defined as 
\begin{equation}
P[\xi ]=N\exp [-\frac{1}{2}\int \int {\xi }\left( x\right) {B^{-1}}\left(
x-y\right) {\xi }\left( y\right) dxdy],
\end{equation}%
where $N$ is the normalization factor. Thus the field $\xi \left( x\right) $
can be regarded as Gaussian random field. Inclusion of higher order terms in
the above makes us possible to treat the non-Gaussian properties as well. On
the other hand, the real part of the action 
\begin{equation}
Re\Gamma +\int \xi \varphi _{\Delta }
\end{equation}%
should be regarded as the standard action which describes dynamics. Applying
the variational principle for the variable $\varphi _{\Delta }(x)$%
\begin{equation}
\left( \frac{\delta \left( Re\Gamma +\int \xi \varphi _{\Delta }\right) }{%
\delta \varphi _{\Delta }(x)}\right) _{\varphi _{\Delta }(x)=0}=-J_{C}(x)
\end{equation}%
we obtain the evolution equation for $\varphi _{C}(x)\equiv \left( \varphi
_{+}(x)+\varphi _{-}(x)\right) /2$, which is simply denoted as $\varphi (x)$
hereafter, 
\begin{equation}
(\partial \partial +m^{2})\varphi +\int_{-\infty }^{t}{d{t}^{\prime }\int {d{%
x}^{\prime }A(x-{x}^{\prime })\varphi }}({x}^{\prime })=\xi \left( x\right) ,
\label{Langevin}
\end{equation}%
where the source term $J_{C}(x)$ is dropped. This is the Langevin equation
for the classical order parameter $\varphi (x)$ with classical random field $%
\xi \left( x\right) $. Actually, if we define the statistical average by%
\begin{equation}
\left\langle {...}\right\rangle _{^{\xi }}\equiv \int {[d\xi ]}...P[\xi ],
\end{equation}%
then the correlation function for the auxiliary fields becomes 
\begin{equation}
\left\langle {\xi (x)\xi (y)}\right\rangle _{\xi }=B(x-y).
\end{equation}%
Moreover, Eq.(\ref{Langevin}) is necessarily causal \cite{Morikawa86} since
the kernel ${{A(x-{x}^{\prime })}}$ has the property ${{A(x-{x}^{\prime })=0}%
}$ for $x^{0}<x^{\prime 0}$. Thus the full effective action is now rewritten
as the statistical superposition of dynamical effective action. Note that
the above dissipative nature arises as a result of an implicit and
systematic coarse graining of the original full theory through the process
of renormalization in the in-in formalism. This renormalization process is
generally inevitable in quantum field theory to obtain finite physical
quantities. The dissipative nature however cannot arise in the ordinary
vacuum as in the standard quantum field theory. For dissipation to arise,
the system must be unstable or permit particle production or be immersed in
the dissipative environment.

The above Langevin equation can, in principle, be transformed into the
equivalent Fokker-Plank equation for the distribution function. This process
is actually tractable if we can use the Markovian property.

The most of the application of the above formalism to the BEC is given in 
\cite{Stoof99}. The Langevin equation is transformed into the Fokker-Plank
equation 
\begin{equation}
i\hbar \frac{\partial }{\partial t}P\left[ \phi ;t\right] =-\sum_{k}\frac{%
\partial }{\partial \phi _{k}}\left( \left( \varepsilon ^{\prime }-iR-\mu
\right) \phi _{k}P\left[ \phi ;t\right] \right) -\frac{1}{2}\sum_{k}\frac{%
\partial ^{2}}{\partial \phi _{k}^{2}}\left( \hbar \Sigma ^{K}P\left[ \phi ;t%
\right] \right) ,
\end{equation}%
for the probability distribution function $P\left[ \phi ;t\right] $, where $%
\varepsilon ^{\prime }$ and $\mu $ are, respectively, renormalized energy
and chemical potential which are related to the function $D$, and $R,$\ $%
\Sigma ^{K}$\ are the transport coefficient functions which are related to
the function $A$ and $B$. This Fokker-Plank equation is further transformed
into the rate equation for the energy density of the condensation $\rho
\left( k;t\right) .$ In the limit of small condensation, i.e.\ the back
reaction of the condensation to the gas is negligible, the rate equation
reduces to 
\begin{equation}
\frac{\partial }{\partial t}\rho \left( k;t\right) \approx \Gamma
^{in}\left( {1-\frac{\rho \left( k;t\right) }{\rho ^{eq}\left( k;t\right) }}%
\right) \approx \Gamma ^{in},  \label{rrateeq}
\end{equation}%
where $\rho ^{eq}\left( k;t\right) $ is the equilibrium value of the
condensation and $\Gamma ^{in}\equiv -R+i\Sigma ^{K}/2$. The order estimate
of the results in \cite{Stoof99} becomes, 
\begin{equation}
\Gamma ^{in}\approx \left\vert \lambda \right\vert ^{2}m^{-3}\rho _{gas}^{2},
\label{Gammain4}
\end{equation}%
if the dominant interaction is $\lambda \left( \phi \phi ^{\ast }\right)
^{2} $, and 
\begin{equation}
\Gamma ^{in}\approx \left\vert \alpha \right\vert ^{2}m\rho _{gas},
\label{Gammain2}
\end{equation}%
if the dominant interaction is of type $\alpha \phi A\partial \phi ^{\ast }$%
, where $A$ is the radiation field. In either case, the induced process is
neglected in the RHS of Eq.(\ref{rrateeq}). It is most interesting that the
detail of the form of $\Gamma ^{in}$ is not relevant for the BEC cosmology (%
\cite{Fukuyama07}). Any choice of $\Gamma ^{in}$ (or $\Gamma $ in the later
argument) yields qualitatively the same cosmology. In this flexible
possibilities, we adopt Eq.(\ref{Gammain4}) in the following although Eq.(%
\ref{Gammain2}) is equally applicable.

The above results determine the BEC condensation rate in the uniform boson
gas. Note that the above condensation rates are transport coefficients which
cannot be found directly in any Lagrangian. This fact yields essential
difference between our approach and the presently popular interacting DE-DM
scalar field models. For example \cite{Amendola2000} \cite{Guo07} study
DE-DM coupling as an interaction term in Lagrangian. The recent reference 
\cite{Gabela09} study DE-DM coupling phenomenologically by setting $w\equiv
p/\rho $ as a constant parameter. The quantity $w$\ actually evolves in time
in our model according to the equation of motion. In these aspects, our BEC
model should be distinguished from general DE-DM coupling models.

So far we have considered the generation of BEC. While this condensation
proceeds, the uniform BEC eventually becomes unstable due to the Jeans
instability and collapses into localized objects (inhomogeneous DM) leaving
no uniform component. If the boson gas has attractive interaction, then this
collapse would be further enhanced.

Thus in the universe, BEC is generated and collapses. These dynamics can be
regarded as generalized phase transitions which cannot be described only by
the coherent time change governed by a Lagrangian. We now examine the detail
of the BEC phase transition in the following subsections.

\subsection{Homogeneous mode: steady BEC}

The gradual generation and subsequent decay of the spatially uniform BEC can
be summarized by the following set of equations in a compact form

\begin{eqnarray}
H^{2} &=&\left( {\frac{\dot{a}}{a}}\right) ^{2}=\frac{8\pi G}{3c^{2}}\left( {%
\rho _{g}+\rho _{\phi }+\rho _{l}}\right) ,  \notag \\
\dot{\rho}_{g} &=&-3H\rho _{g}-\Gamma \rho _{g},  \notag \\
\dot{\rho}_{\phi } &=&-6H\left( {\rho _{\phi }-V}\right) +\Gamma \rho _{g}-{%
\Gamma }^{\prime }\rho _{\phi },  \notag \\
\dot{\rho}_{l} &=&-3H\rho _{l}+{\Gamma }^{\prime }\rho _{\phi }.
\label{EqsetforBECcosmology}
\end{eqnarray}%
where we assume a simple form of potential $V\equiv V_{0}+m^{2}\phi ^{\ast
}\phi +\frac{\lambda }{2}(\phi ^{\ast }\phi )^{2}$ with attractive
interaction $\lambda <0$. This potential form will be fully fixed if we can
identify the boson field in the future. The Einstein equation in the first
line of Eq.(\ref{EqsetforBECcosmology}) gives the cosmic expansion rate $H$
in terms of the energy densities of gas $\rho _{g}$, BEC $\rho _{\phi }$,
and the localized objects $\rho _{l}$. The second line represents the
evolution of the gas energy density $\rho _{g}$\ which reduces due to the
cosmic expansion $-3H\rho _{g}$ and the consumption for BEC $-\Gamma \rho
_{g}.$ The third line is the relativistic form of Gross-Pitaevski equation
describing BEC dynamics with the steady condensation rate $\Gamma \rho _{g}$%
, where $\Gamma =const.$ (for Eq.(\ref{Gammain2})) or $\Gamma \propto \rho
_{g}$ (for Eq.(\ref{Gammain4})). This form of source and loss terms $\pm
\Gamma \rho _{g}$ is the most characteristic to our BEC model although they
are small. In the same line of Eq.(\ref{EqsetforBECcosmology}), the term $-{%
\Gamma }^{\prime }\rho _{\phi }$\ symbolically represents a sudden collapse
of BEC. The detail of this collapse will be explained in the next
subsection. The last line represents the evolution of the localized energy
density $\rho _{l}$\ which reduces due to the cosmic expansion $-3H\rho _{l}$
and increases due to the BEC collapse $+{\Gamma }^{\prime }\rho _{\phi }$.
We now explain the BEC instability and the collapse.

\subsection{Inhomogeneous modes: BEC collapse}

Uniform BEC is not an absolutely stable phase if the interaction is
attractive or gravity is concerned. It collapses into localized object after
the appearance of instability. Let us estimate the scale of BEC instability,
which determines the time scale of the BEC collapse and the resultant mass
scale of the localized objects. We start the space-time metric of the form%
\begin{equation}
\mathrm{d}s^{2}=(1+2\Phi )\mathrm{d}t^{2}-a^{2}(1-2\Phi )\mathrm{d}\mathbf{x}%
^{2},  \label{eqn:metric2}
\end{equation}%
where $\Phi =\Phi \left( t,\vec{x}\right) $ represents the gravitational
potential and $a=a\left( t\right) $ the cosmic scale factor. BEC is
described by the equation of motion for the order parameter $\phi =\phi
\left( t,\vec{x}\right) $, i.e. the third line of Eq.(\ref%
{EqsetforBECcosmology}) neglecting ${\Gamma }$ and ${\Gamma }^{\prime }$
since the instability time scale is very short compared with $\Gamma ^{-1}$
and $\Gamma ^{\prime -1}$. Up to the first order in $\Phi $, this equation
becomes 
\begin{equation}
\ddot{\phi}+3\frac{\dot{a}}{a}\dot{\phi}-\frac{1}{a^{2}}\nabla ^{2}\phi
+m^{2}(1+2\Phi )\phi +\lambda (1+2\Phi )|\phi |^{2}\phi =0\,.
\label{eqn:scalar}
\end{equation}%
Gravity is described by the Poisson equation, in the Einstein equation, 
\begin{equation}
\nabla ^{2}\Phi =4\pi Ga^{2}\left\{ \dot{\phi}^{\dagger }\dot{\phi}+\frac{%
\nabla }{a}\phi \cdot \frac{\nabla }{a}\phi ^{\dagger }+m^{2}|\phi |^{2}+%
\frac{\lambda }{2}|\phi |^{4}-\rho _{0}\right\} ,  \label{eqn:Poisson}
\end{equation}%
where $\rho _{0}$ is the uniform background energy density. We consider the
linear perturbation on the uniform background

\begin{equation}
\phi \left( t,\vec{x}\right) =\phi _{0}\left( t\right) +\phi _{1}\left( t,%
\vec{x}\right) \,,\Phi \left( t,\vec{x}\right) =0+\Phi _{1}\left( t,\vec{x}%
\right) .
\end{equation}%
where small perturbations $\phi _{1}$ and $\Phi _{1}$ are decomposed into
Fourier modes, 
\begin{eqnarray}
\phi _{1} &=&\phi _{1}\exp (\Omega t+i\overrightarrow{k}\overrightarrow{r}),
\notag \\
\Phi _{1} &=&\Phi _{1}\exp (\Omega t+i\overrightarrow{k}\overrightarrow{r}).
\end{eqnarray}%
The background variables satisfy, from Eqs.(\ref{eqn:scalar}-\ref%
{eqn:Poisson}), 
\begin{eqnarray}
\ddot{\phi}_{0}+3\frac{\dot{a}}{a}\dot{\phi}_{0}+\left( m^{2}+\lambda |\phi
_{0}|^{2}\right) \phi _{0} &=&0\,,  \label{eqn:scalar-0th} \\
\dot{\phi}_{0}^{\dagger }\dot{\phi}_{0}+\left( m^{2}+\frac{\lambda }{4}|\phi
_{0}|^{2}\right) |\phi _{0}|^{2}-\rho _{0} &=&0.  \label{eqn:Poisson-0th}
\end{eqnarray}%
In these Eqs.(\ref{eqn:scalar-0th}-\ref{eqn:Poisson-0th}), the variable $%
\phi _{0}$ can be assumed to be real without generality since all the
coefficients are real. Then the existence of a non-trivial solution requires
that the above set of linear equations are dependent with each other. Thus
we have the condition 
\begin{eqnarray}
&&\left[ \left( \Omega ^{2}+3\frac{\dot{a}}{a}\Omega +\frac{\mathbf{k}^{2}}{%
a^{2}}+m^{2}+3\lambda \phi _{0}^{2}\right) \frac{\mathbf{k}^{2}}{4\pi Ga^{2}}%
\right.  \notag  \label{det} \\
&&\left. ~~~~~~-2(m^{2}+\lambda \phi _{0}^{2})\phi _{0}\left( 2\dot{\phi}%
_{0}\Omega +2(m^{2}+\lambda \phi _{0}^{2})\phi _{0}\right) \right]  \notag \\
&\times &\left( \Omega ^{2}+3\frac{\dot{a}}{a}\Omega +\frac{\mathbf{k}^{2}}{%
a^{2}}+m^{2}+\lambda \phi _{0}^{2}\right) =0\,,  \label{instability}
\end{eqnarray}%
which determines the instability strength $\Omega $ as a function of the
wave number $\vec{k}.$ The wave number $\vec{k}$ which corresponds to the
maximum value of $\Omega $ (\textsl{i.e.} most unstable mode) is not
necessarily the relevant scale since we also have to consider the completion
time of the collapse $a/\left( \alpha k\right) $, where $\alpha $ $(<1)$ is
the decay speed of BEC. Thus the relevant scale will be the wave number $%
k_{\ast }$ which should satisfy the condition that the collapse time is just
the instability time i.e. $\alpha k_{\ast }/a=\Omega .$ Choosing the value
associated with the most unstable mode among four solutions of $\Omega $, we
have 
\begin{equation}
\frac{k_{\ast }}{a}=\left( \frac{-m_{eff}^{2}+\sqrt{m_{eff}^{4}+64\pi
G(1+\alpha ^{2})\left( m^{4}\phi _{0}^{2}-2\kappa m^{2}\phi _{0}^{4}+\kappa
^{2}\phi _{0}^{6}\right) }}{2(1+\alpha ^{2})}\right) ^{1/2}
\label{solk=omega}
\end{equation}%
where $m_{eff}^{2}\equiv m^{2}-3\kappa \phi _{0}^{2}$, $\kappa =-\lambda
(>0) $ and adiabatic approximation $H=0,$\ $\dot{\phi}_{0}=0$\ is utilized
since the collapse time scale is much smaller than cosmic and condensation
time scales in the late universe.

The actual scale critically depends on the boson mass\cite{Fukuyama07}. If
it is about $1$eV, then the typical mass of the structure will be $M_{\ast
}\approx 1.6\times 10^{11}M_{\odot }$ which is of order of a galaxy. If it
is about $10^{-3}$eV, then the typical scale in tiny $M_{\ast }\approx
10^{-1}g$ and therefore BEC collapse cannot form macroscopic cosmological
structures. In any case, these collapsed clumps would work as the ordinary
DM and the scenario of large scale structure formation exactly reduces to
the standard model.

A typical numerical demonstration of uniform BEC mode is given in Fig.(\ref%
{fig1}). Most of the characteristic features of BEC cosmology are shown in
this figure. The gas density $\rho _{g}$\ (grey line) reduces monotonically
while the BEC density $\rho _{\phi }$\ (solid line) gradually increases
toward some instability point where it suddenly decays into localized
objects $\rho _{l}$ (dashed line). Then the BEC density disappears and a new
BEC begins. This cycle repeats several times and finally the phase of $\rho
_{\phi }=const.$ appears. Thus the universe inevitably enters into the final
accelerated expansion phase. This phase is guaranteed by the stable balance
of the condensation flow and the potential force, and is an attractor of Eq.(%
\ref{EqsetforBECcosmology}). Several other features and predictions of this
model are found in \cite{Fukuyama07}.

\begin{figure}[h]
\begin{center}
\begin{tabular}{cc}
\resizebox{120mm}{!}{\includegraphics{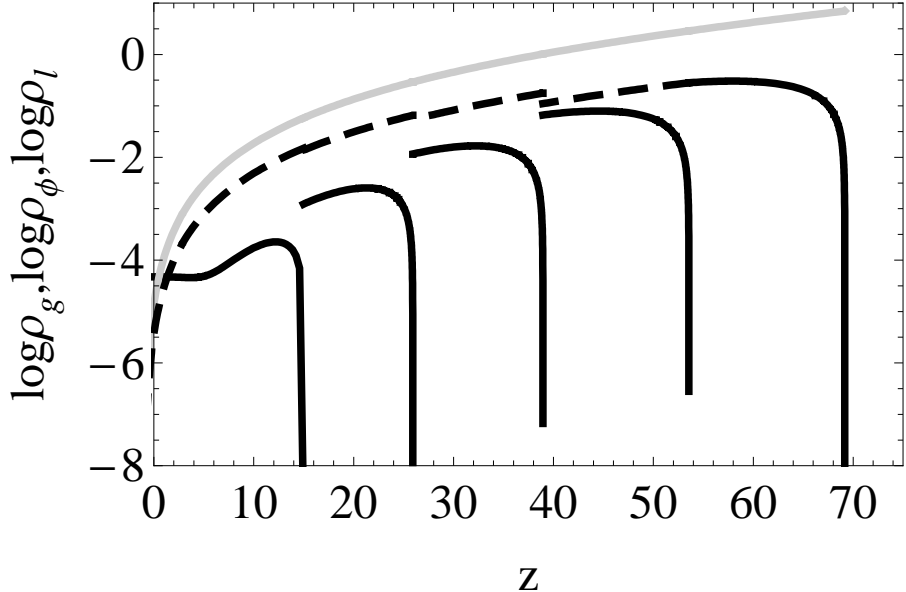}} &  \\ 
& 
\end{tabular}%
\end{center}
\caption{A typical numerical demonstration of the BEC cosmology. This graph
represents the evolution of energy densities of the boson gas $\protect\rho %
_{g}$ (grey line), BEC $\protect\rho _{\protect\phi }$ (solid line), and the
localized objects $\protect\rho _{l}$ (dashed line) as a function of
redshift $z$. We solved Eq.(\protect\ref{EqsetforBECcosmology}) with the
parameters $\protect\lambda =-0.51$, $\Gamma =0.2601$, $m=0.0028eV$. The BEC
Collpase time sequence is $z=69,54,39,26,15$. After four collapses of BEC,
the energy density of BEC $\protect\rho _{\protect\phi }$ becomes constant,
which induces the accelerated expansion of the present universe. }
\label{fig1}
\end{figure}

\section{BEC in the early universe}

We now extend the BEC cosmology to the early universe. As was studied
previously, for the relativistic case, the critical temperature $%
T_{cr}\propto \rho ^{1/2}$ increases much faster than the cosmic temperature 
$T\propto \rho ^{1/4}$ in the course of increasing energy density $\rho $.
Therefore we safely expect the universe is always under critical temperature
provided so at present.

The acceleration in the early universe is quite different from that in late
universe in the sense that a huge vacuum energy $V_{0}$ is present from the
beginning in the former. However we do not want to introduce any new
ingredients into the model used in the original analysis in LA. This is
because we want to understand the cosmic accelerations from a unified point
of view. Actually, the mass $m$ of the boson is about eV and therefore is
completely negligible in the early universe. The condensation rate $\Gamma $%
\ has been quite robust in the previous study in the contemporary universe
and therefore we take the same form of $\Gamma $ for BEC condensation as
used previously.

The goal of this section is to examine the following individual steps to
complete the whole BEC cosmological scenario and verify this model: 1.
Natural initiation of the inflation from the fireball stage is expected from
the steady BEC processes. 2. This inflation stage terminates autonomously
due to the stagflation. 3. Reheating process inevitably takes place due to
the non-linear coupling in BEC. 4. The cosmological constant is autonomously
adjusted to be zero due to the BEC instability at the stagflation point. 5.
We calculate the generated density fluctuations and compare them with
observations.

\subsection{Natural initiation of the inflation}

The original model of the BEC cosmology is described by Eq.(\ref%
{EqsetforBECcosmology}). Now in the application to the early universe whose
energy scale is extremely higher than eV, a typical mass scale of the boson,
relativistic damping terms should be used:

\begin{eqnarray}
H^{2} &=&\left( {\frac{\dot{a}}{a}}\right) ^{2}=\mu ^{2}\left( {\rho
_{g}+\rho _{\phi }+\rho _{l}}\right) ,  \notag \\
\dot{\rho}_{g} &=&-4H\rho _{g}-\Gamma \rho _{g},  \notag \\
\dot{\rho}_{\phi } &=&-6H\left( {\rho _{\phi }-V}\right) +\Gamma \rho _{g}-{%
\Gamma }^{\prime }\rho _{\phi },  \notag \\
\dot{\rho}_{l} &=&-4H\rho _{l}+{\Gamma }^{\prime }\rho _{\phi }.
\label{eqn:BECcosmologyRel}
\end{eqnarray}%
where $\mu ^{2}=8\pi G/(3c^{2})$, the potential is $V\equiv V_{0}+m^{2}\phi
^{\ast }\phi +\frac{\lambda }{2}(\phi ^{\ast }\phi )^{2}$ with $\lambda <0,$
and the energy density for BEC is $\rho _{\phi }=\dot{\phi}^{\ast }\dot{\phi}%
+V.$

In the early universe, tiny mass term $m^{2}$ is neglected. On the other
hand, the condensation rate $\Gamma $, whose energy scale dependence is not
clear, may still be relevant in the early universe. We should consider an
initial vacuum energy $V_{0}\approx m_{\text{pl}}^{4}$, which is huge up to $%
10^{120}$ times the present value.

Initial boson gas $\rho _{g}\left( t\right) =\rho _{gi}\left( t_{i}/t\right)
^{2}$\ and the temperature will be reduced by the cosmic expansion $%
a(t)=a_{0}\left( t/t_{i}\right) ^{1/2}$. On this background, BEC gradually
proceeds since $T_{cr}<T$ is guaranteed from the energy density dependence
of them as argued above. Since the initial evolution velocity of the BEC $%
\dot{\phi}$ is small and the potential energy is negligible compared with
the boson gas density, the BEC $\phi $\ approximately obeys 
\begin{equation}
3H\dot{\phi}^{2}=\Gamma \rho _{g},
\end{equation}%
whose solution, under the choice $\Gamma \left( t\right) =\Gamma _{0}\rho
_{g}\left( t\right) $, is given by the fire ball solution, 
\begin{equation}
\phi _{\text{fire}}\left( t\right) =2t_{i}\left( \frac{\Gamma _{0}\rho
_{gi}^{3/2}}{3\mu }\right) ^{1/2}\left( 1-\left( \frac{t_{i}}{t}\right)
^{1/2}\right) .  \label{phi-fire}
\end{equation}%
The situation of each stage of BEC will be given in Fig.(\ref{fig4}). Then,
if we identify the initiation time of inflation t$_{\text{iinf}}$\ by the
condition 
\begin{equation}
\rho _{g}\left( t_{\text{iinf}}\right) =V_{0},  \label{match-phi-fire-inf}
\end{equation}%
then the initial value of the BEC when the inflation begins will be 
\begin{equation}
\phi _{\text{fire}}\left( t_{\text{iinf}}\right) =2t_{i}\left( \frac{\Gamma
_{0}\rho _{gi}^{3/2}}{3\mu }\right) ^{1/2}\left( 1-\left( \frac{V_{0}}{\rho
_{gi}}\right) ^{1/4}\right)   \label{phi-fire-iinf}
\end{equation}%
or 
\begin{equation}
\delta \equiv \dot{\phi}_{\text{fire}}\left( t_{\text{iinf}}\right) =\left( 
\frac{\Gamma _{0}\rho _{gi}^{3/2}}{3\mu }\right) ^{1/2}\left( \frac{V_{0}}{%
\rho _{gi}}\right) ^{3/4}.  \label{delta}
\end{equation}%
\newline

Although the boson gas will be diluted, this initial BEC triggers the
further evolution of BEC, and BEC proceeds consuming the vacuum energy.
During the inflation, BEC evolves as 
\begin{equation}
\dot{V}=-3H\dot{\phi}^{2},\;H^{2}=\mu ^{2}V  \label{app-inflation}
\end{equation}%
which solves, under the initial condition Eq.(\ref{delta}), as%
\begin{equation}
\phi _{\text{inf}}\left( t\right) =\left( \frac{3H_{\text{inf}}}{-2\lambda }%
\right) ^{1/2}\left( t_{\text{iinf}}-t+\frac{1}{2}\left( \frac{3H_{\text{inf}%
}}{-\lambda \delta ^{2}}\right) ^{1/3}\right) ^{-1/2},  \label{phi-inf}
\end{equation}%
where we approximate $H^{2}=H_{\text{inf}}^{2}\equiv \mu ^{2}V_{0}$. Since
the potential is flat around the origin, this solution actually satisfies
the slow-role condition 
\begin{equation}
\left\vert \frac{\dot{\phi}_{\text{inf}}^{2}}{2V}\right\vert \ll
1,\left\vert \frac{\ddot{\phi}_{\text{inf}}}{3H_{\text{inf}}\dot{\phi}_{%
\text{inf}}}\right\vert \ll 1
\end{equation}%
or equivalently under the condition Eq.(\ref{app-inflation}), 
\begin{equation}
\left\vert \varepsilon \right\vert \ll 1,\;\left\vert \eta \right\vert \ll 1,
\end{equation}%
where 
\begin{equation}
\varepsilon \equiv \frac{m_{pl}^{2}}{16\pi }\left( \frac{V^{\prime }}{V}%
\right) ^{2},\;\eta \equiv \frac{m_{pl}^{2}}{8\pi }\frac{V^{\prime \prime }}{%
V}.  \label{epsilon and eta}
\end{equation}%
These conditions hold until just before $\phi _{\text{inf}}\left( t\right) $
hits the (virtual) singularity at $t=t_{\text{einf}}$ (where $\phi _{\text{%
inf}}$ diverges). Thus the duration of the inflation is%
\begin{equation}
\Delta _{\text{inf}}\equiv t_{\text{einf}}-t_{\text{iinf}}=\frac{1}{2}\left( 
\frac{3H_{\text{inf}}}{-\lambda \delta ^{2}}\right) ^{1/3}+O\left( H_{\inf
}^{-1}\right)  \label{Delta-inf}
\end{equation}%
and the e-folding number $N$ becomes $N=H_{\text{inf}}\Delta _{\text{inf}}$.

\begin{figure}[h]
\begin{center}
\begin{tabular}{cc}
\resizebox{120mm}{!}{\includegraphics{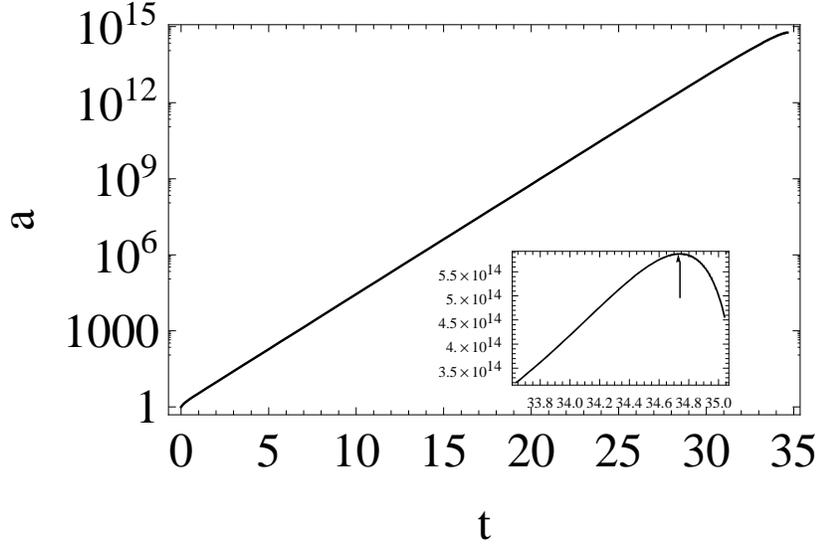}} &  \\ 
& 
\end{tabular}%
\end{center}
\caption{The time evolution of the scale factor $a\left( t\right) .$ The
scale factor is normalized by itself at the begining of the inflation. This
is a solution of Eq.(\protect\ref{eqn:BECcosmologyRel}) with parameters: $%
\protect\lambda =-0.5,V_{0}=1,$ $\protect\phi \left( 0\right) =0,$ $\dot{%
\protect\phi}\left( 0\right) =0.001,.\protect\rho _{g}\left( 0\right)
=5,\Gamma _{0}=1,$ in the units of $H_{\inf }$ for $H$, $H_{\inf }^{-1}$ for 
$t$, $H_{\inf }^{2}/\protect\mu ^{2}$ for $\protect\rho $, where $H_{\inf
}^{2}\equiv \protect\mu ^{2}V_{0}$. The inflationary phase, i.e. $a\propto
e^{H_{\inf }t},$ naturally appears with sufficient e-folding number. The
total energy density vanishes and the expansion ceases at the stagflation
point $t=t_{s}\approx 34.8$, where the strong instability of the uniform BEC
mode actually invalidates the evolution after $t_{s}$. The inset magnifies
the local region around $t_{s},$ which is marked by an arrow. }
\label{fig2}
\end{figure}

In Fig.(\ref{fig2}), we have depicted a typical time evolution of the scale
factor. We do not seek for the most favorable values of parameters in this
paper since our aim is to clarify the basic mechanism and to show the
robustness of the BEC cosmological model in the early universe.

This mechanism of inflation is the same as the standard models though the
initiation of it is quite different. Moreover, the termination of it is also
different as we will see in the next subsection. In the inflationary phase,
BEC develops by consuming the vacuum energy contrary to the case of BEC
evolution in LA in which BEC develops by consuming the boson gas energy.

\subsection{Termination of inflation}

As is shown in Fig.(\ref{fig2}), this inflation sharply terminates within a
finite time. This can be easily understood as a general feature of our BEC
cosmology. In the late stage of inflation $\Gamma ,\Gamma ^{\prime },\rho
_{g}$ are neglected, and Eq.(\ref{eqn:BECcosmologyRel}) reduces to 
\begin{equation}
H^{2}=\frac{8\pi G}{3c^{2}}{\rho _{\phi }},\dot{\rho}_{\phi }=-3H\dot{\phi}%
^{2},  \label{redset}
\end{equation}%
and, provided $H\neq 0$, further to

\begin{equation}
\dot{H}=-\frac{3}{2}\mu ^{2}\dot{\phi}^{2}.  \label{Hdot}
\end{equation}%
Since the right hand side of Eq.(\ref{Hdot}) is negative and $\dot{\phi}^{2}$
increases in time, the expansion rate $H$ eventually crosses 0, at time $%
t=t_{s}$, after that the universe terns into the contraction phase. The
dynamics of BEC in this process corresponds to the over-hill regime in the
original BEC cosmology \cite{Fukuyama06}.

When the cosmic expansion ceases at $t=t_{s}$, Eq.(\ref{eqn:BECcosmologyRel}%
) yields, neglecting $\rho _{g},$ 
\begin{equation}
H=0,{\rho _{\phi }=0},\dot{\rho}_{\phi }=0,  \label{stagflation}
\end{equation}%
which guarantee the energy density smoothly vanishes at $t=t_{s}$.\ However,
we should notice that $\dot{H}\neq 0$ from Eq.(\ref{Hdot}) and the
continuity. We call this phase as \textsl{stagflation }regime, i.e. the
cosmic expansion is 'stagnant' while the BEC is steadily accelerating and
'inflating'. In the previous approximate solution Eq.(\ref{phi-inf}), the
termination time $t_{\text{einf}}$ is almost $t_{s}$: $t_{s}\approx t_{\text{%
einf}}$.

\begin{figure}[h]
\begin{center}
\begin{tabular}{cc}
\resizebox{120mm}{!}{\includegraphics{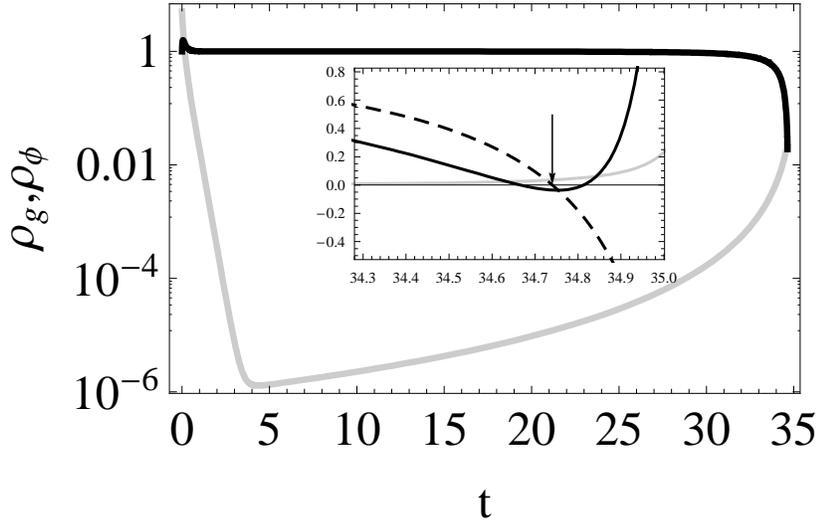}} &  \\ 
& 
\end{tabular}%
\end{center}
\caption{This graph represents the evolution of energy densities of the
boson gas $\protect\rho _{g}$ (grey line), and of BEC $\protect\rho _{%
\protect\phi }$ (solid line) as a function of time $t$. The enrgy density is
normalized by $\left( H_{\inf }/\protect\mu \right) ^{2}$ and the time by $%
H_{\inf }^{-1}$. The inflationary phase eventually terminates and the
universe turns into the contraction phase. It finally approaches the virtual
singularity, if we do not consider the instability and the decay of BEC.
Inset is the magnified graph at the last stage of the inflation. Dashed line
is the Hubble parameter, which crosses zero and the total energy density
vanishes at the stagflation point $t=t_{s}\approx 34.74$ (marked by an
arrow). The strong instability of the uniform BEC mode actually invalidates
the evolution after $t_{s}$.}
\label{fig3}
\end{figure}

Figure (\ref{fig3}) shows the energy densities of the boson gas $\rho _{g}$
(grey line), and of BEC $\rho _{\phi }$ (solid line) in the same time
evolution as in Fig.(\ref{fig2}). The stagflation regime is magnified in the
inset of this figure. The dashed line shows the Hubble parameter which
smoothly changes its signature, expanding to contracting at $t_{s}$.

\begin{figure}[h]
\begin{center}
\begin{tabular}{cc}
\resizebox{120mm}{!}{\includegraphics{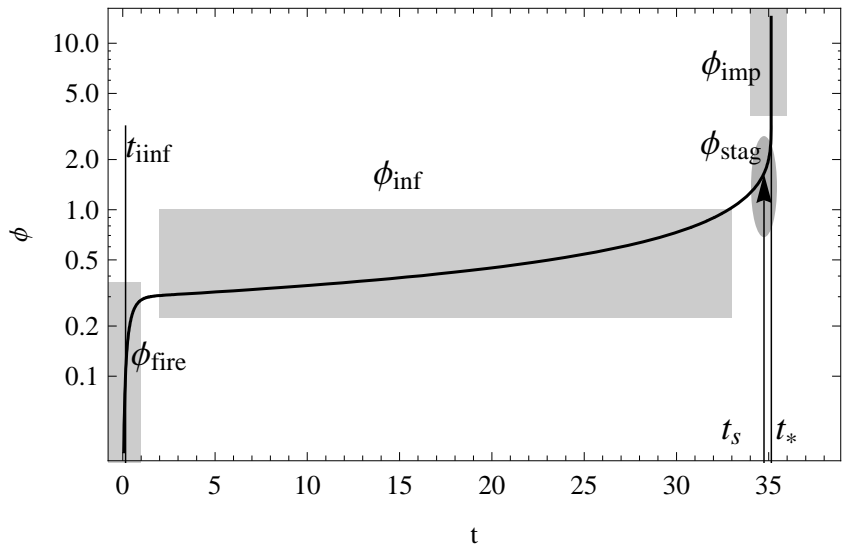}} &  \\ 
& 
\end{tabular}%
\end{center}
\caption{This figure represents the evolution of BEC field $\protect\phi %
\left( t\right) $, numerically obtained, and several analytical
approximations, $\protect\phi _{\text{fire}}$, $\protect\phi _{\text{inf}}$, 
$\protect\phi _{\text{stag}}$, and $\protect\phi _{\text{imp}}$. The
applicability range for each solution is shown by shaded region.The
solutions $\protect\phi _{\text{fire}}$, $\protect\phi _{\text{inf}}$ are
matched together at $t_{\text{i}\inf }\approx 0.16$. The stagflation time $%
t_{s}\approx 34.7$ is marked by an arrow. The BEC\ field $\protect\phi $\
diverges at the implosion time $t_{\ast }\approx 35.1$. }
\label{fig4}
\end{figure}

We would like to comment briefly on the virtual history of the universe
after the stagflation regime though it actually does not exist, as will be
explained in the following subsections. The stagflation regime Eq.(\ref%
{stagflation}) does not last forever; the kinetic term $\dot{\phi}^{2}$
eventually develops and surpasses $\lambda \phi ^{4}$ term. In this kinetic
term-dominant regime, Eq.(\ref{eqn:BECcosmologyRel}) yields $d\left( \dot{%
\phi}^{2}/2\right) /dt=-3H\dot{\phi}^{2}$ and $H^{2}=\mu ^{2}\dot{\phi}^{2}/2
$, and therefore 
\begin{equation}
\ddot{\phi}=\frac{3}{\sqrt{2}}\mu \dot{\phi}^{2},
\end{equation}%
where we must choose $H<0$ and $\dot{\phi}>0$ for consistent solution. These
lead to the implosive solution%
\begin{equation}
\phi _{\text{imp}}=\text{const.}-\frac{\sqrt{2}}{3\mu }\log \left( t_{\ast
}-t\right) ,  \label{phi-imp}
\end{equation}%
where the universe hits the singularity at time $t_{\ast }$. The cosmic
contraction becomes 
\begin{equation}
a=\text{const.}\left( t_{\ast }-t\right) ^{1/3},
\end{equation}%
which explains the virtual behavior in Fig.(\ref{fig2}) for $t\rightarrow
t_{\ast }.$(See also Fig.(\ref{fig4}).)

As a whole, the early universe suffers the BEC phase transition as depicted
in Fig.(\ref{fig4}). This figure represents the whole evolution of BEC field 
$\phi \left( t\right) $, numerically obtained, and several analytical
approximations whose validity is limited within each local region. Analytic
solutions are $\phi _{\text{fire}}$ in Eq.(\ref{phi-fire}), $\phi _{\text{inf%
}}$ in Eq.(\ref{phi-inf}), $\phi _{\text{stag}}$ in Eq.(\ref{phi-stag})
(which will be described shortly), and $\phi _{\text{imp}}$ in Eq.(\ref%
{phi-imp}). The solutions $\phi _{\text{fire}}$ and $\phi _{\text{inf}}$ are
matched at $t_{\text{i}\inf }$,\ defined by Eq.(\ref{match-phi-fire-inf}).
The stagflation time $t_{s}$, marked by an arrow, is defined by Eq.(\ref%
{stagflation}). The implosion time $t_{\ast }$ is the time when BEC field
diverges as Eq.(\ref{phi-imp}).

In the real world however, the infinite amplitude $\phi \rightarrow \infty $
and velocity $\dot{\phi}\rightarrow \infty $ for BEC are not allowed.
Actually, before the universe enters into this phase, there appear two new
features in this scenario and the singularity, which we call \textsl{%
implosion}, is avoided. One new feature is the particle production due to
the ever accelerating BEC $\dot{\phi}\rightarrow \infty $. This is the
origin of the reheating that some portion of the original vacuum energy $%
V_{0}$ is converted into thermalized gas of bosons. Another new feature is
the instability of the uniform mode of BEC at $t\approx t_{s}$ and $\rho
_{\phi }\approx 0$, and subsequent decay of the uniform BEC into the
localized objects, without leaving any vacuum energy. This is the mechanism
which did adjust the final vanishing vacuum energy, or vanishing
cosmological 'constant', despite the huge vacuum energy set before
inflation. Let us examine each of these features separately in the following.

\subsection{Initiation of reheating}

\ The rapid increase of BEC $\dot{\phi}\rightarrow \infty $ at the late
stage of the inflation yields plenty of particle production, \textsl{which
will reheat the universe}, through the self-interaction. If this particle
production is effective to turn sufficient amount of kinetic energy of BEC
into thermal gas, the universe can go into the big bang phase. In order to
demonstrate this scenario, we consider a simple model of particle
production; an addition of the term $-\gamma \dot{\phi}^{2}$ $(+\gamma \dot{%
\phi}^{2})$ to the right hand side of $\dot{\rho}_{\phi }$ $(\dot{\rho}_{g}$%
, respectively$)$ in Eq.(\ref{eqn:BECcosmologyRel}). We can approximately
evaluate the particle production during the inflationary era. By using Eq.(%
\ref{phi-inf}), we obtain 
\begin{equation}
\rho _{g}\left( t_{0}\right) =\gamma \int_{t_{\text{iinf}}}^{t_{0}}\dot{\phi}%
_{\text{inf}}^{2}dt\leq \frac{\gamma V_{0}}{3H_{\text{inf}}}=\frac{\gamma H_{%
\text{inf}}}{3\mu ^{2}}.
\end{equation}%
If we set $\gamma \approx m_{\phi }\approx 1eV$, then the reheating
temperature, using the expression $\rho _{g}=\left( \frac{\pi ^{2}g_{\ast
}k_{B}^{4}}{30\hbar ^{3}c^{3}}\right) T^{4}$, would be 
\begin{equation}
k_{B}T_{rh}<\left( \frac{10}{\pi ^{2}}\gamma m_{pl}^{3}\right) ^{1/4}\left(
H_{\inf }/m_{\phi }\right) ^{1/4}\approx 1.17\times 10^{12}\left( H_{\inf
}/m_{pl}\right) ^{1/4}GeV.
\end{equation}

Note that the particle production in the BEC model is caused simply by the
accelerated evolution of the background field $\phi $.

\subsection{Autonomous adjustment toward the vanishing cosmological constant}

Let us discuss on another new feature which does not allow the implosive
singularity. As in the previous arguments on the BEC cosmology in LA, the
instability of BEC actually develops and the homogeneous BEC phase collapses
into many localized objects. Thus dynamics of BEC is not closed within the
homogeneous mode contrary to the recent arguments on the DE-DM coupling (%
\cite{Guo07}) for example.

In the middle of the inflationary phase, the BEC fluctuations cannot
develop. However, they have a chance to grow after the inflationary phase.
We now examine this around the stagflation time when the energy density
vanishes at $t=t_{s}.$\ 

The middle of Eq.(\ref{stagflation}) allows us to approximate the evolution
equation, neglecting $V_{0},$ as

\begin{equation}
\dot{\phi}=\sqrt{\kappa }\phi ^{2},  \label{eq5}
\end{equation}%
which immediately solves 
\begin{equation}
\phi _{\text{stag}}=\kappa ^{-1/2}\left( t_{s}-t\right) ^{-1},
\label{phi-stag}
\end{equation}%
where $\kappa =-\lambda (>0)$. In this regime $t\approx t_{s},$ the
instability scale Eq.(\ref{solk=omega}) becomes

\begin{equation}
\frac{k_{\ast }}{a}=\left( 8\pi \right) ^{1/4}G^{1/4}\kappa ^{1/2}\phi _{%
\text{stag}}^{3/2}=\left( 8\pi \right) ^{1/4}G^{1/4}\kappa ^{-1}\left(
t_{s}-t\right) ^{-3/2},  \label{eq7}
\end{equation}%
which increases without bound for $t\rightarrow t_{s}$. On the other hand,
the cosmic expansion is almost ceased in this regime $t\approx t_{s}$.
Therefore, we expect a strong instability takes place and BEC rapidly
collapses into localized objects, whose density is denoted as $\rho _{l}$.
Moreover, there is sufficient time $\left( t_{s}-t\right) $\ for the
completion of the BEC collapse before the universe escapes from this regime $%
t\approx t_{s}$: 
\begin{equation}
\left( t_{s}-t\right) >a/k_{\ast }\propto \left( t_{s}-t\right) ^{3/2}.
\label{eq8}
\end{equation}%
This is similar to what happens in the over-hill regime in the late time BEC
cosmology. Thus all the BEC component turns into localized objects, which
behave as dust fluid, leaving vanishing vacuum energy or cosmological
'constant', despite the initial value of the vacuum energy $V_{0}$. The fate
of these localized objects is not yet clear. If they form light blackholes,
they may eventually evaporate into radiation including the boson gas. If
they form heavy blackholes, they may remain until now and yield significant
effects on the cosmic structure. We simply choose the former case in our
numerical demonstration in this paper.

\begin{figure}[h]
\begin{center}
\begin{tabular}{cc}
\resizebox{120mm}{!}{\includegraphics{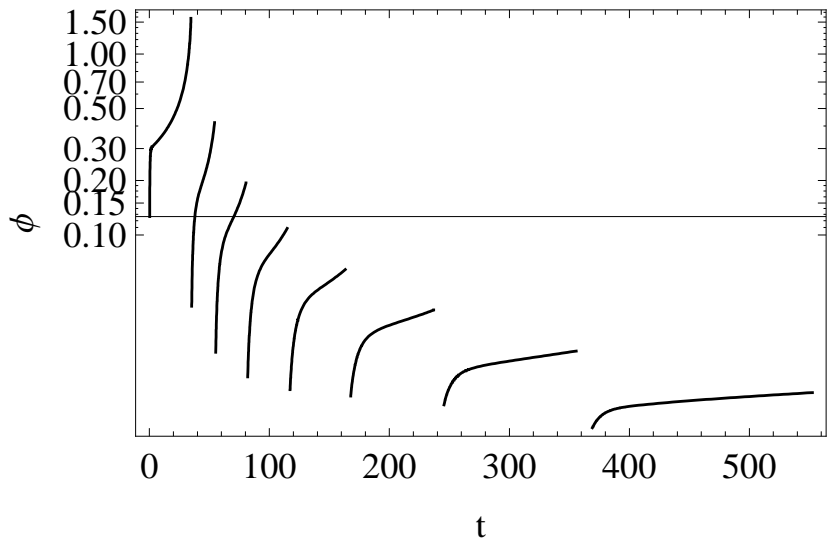}} &  \\ 
& 
\end{tabular}%
\end{center}
\caption{Time evolution of the BEC order parameter $\protect\phi $ is shown.
It shows a sequence of BEC collapses. It may happen that this BEC reduction
sequence leads to the era of LA in Ref.[\protect\ref{fig1}]. However, the
numerical calculation extending over 100 digits is almost impossible and we
could not reveal the full extention of the reduction sequence.}
\label{fig5}
\end{figure}

\begin{figure}[h]
\begin{center}
\begin{tabular}{cc}
\resizebox{120mm}{!}{\includegraphics{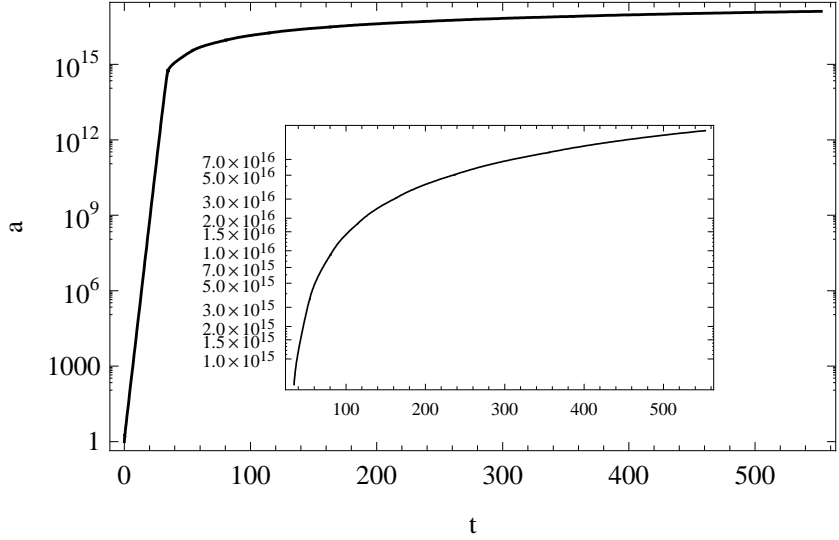}} &  \\ 
& 
\end{tabular}%
\end{center}
\caption{Time evolution of the scale factor $a\left( t\right) $ is shown in
logarithmic scale against the cosmic time $t$. Initial inflation era is
followed by the almost radiation dominated era. The inset is the later
evolution.}
\label{fig6}
\end{figure}

\begin{figure}[h]
\begin{center}
\begin{tabular}{cc}
\resizebox{120mm}{!}{\includegraphics{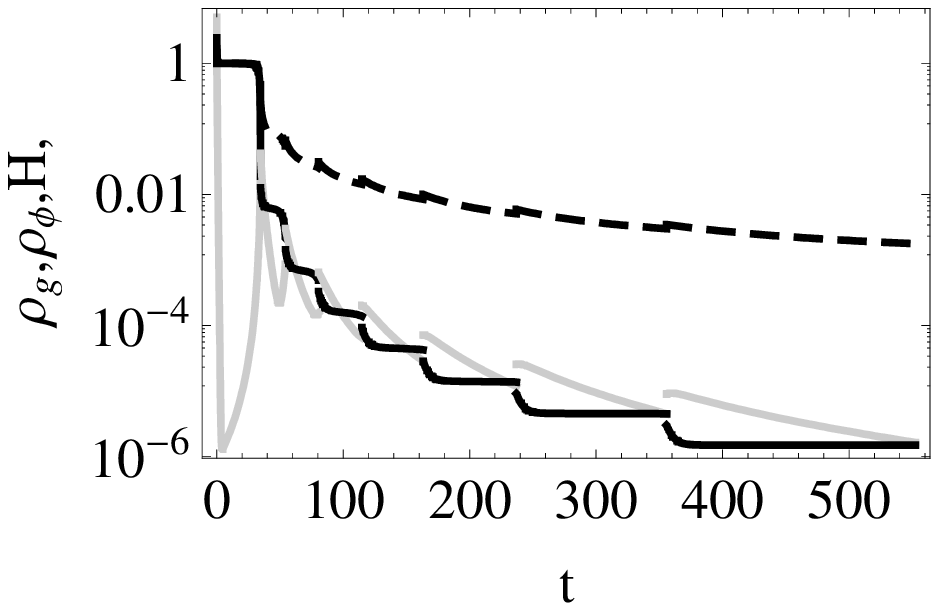}} &  \\ 
& 
\end{tabular}%
\end{center}
\caption{Time evolution of BEC energy density $\protect\rho _{\protect\phi }$%
(solid line), boson gas energy density $\protect\rho _{g}$ (gray line), and
the Hubble parameter $H$ (dashed line) in cosmic time $t$. The original gas
density $\protect\rho _{g}$ reduces almost completely due to the exponential
expansion during the inflationary era. In the later stage of the
inflationary era, uniform BEC becomes unstable and collapses leaving some
fraction of $\protect\rho _{\protect\phi }$. At the same time, $\protect\rho %
_{g}$ gradually recovers due to the particle production process caused by
the rapid change of the background field $\protect\phi $ and evetually
dominates $\protect\rho _{\protect\phi }$. However $\protect\rho _{g}$
reduces faster than $\protect\rho _{\protect\phi }$ in the subsequent cosmic
evolution. Eventually the second instability of BEC takes place. These
sequence of BEC and its decay repeats and the effective csmological constant
reduces toward zero. }
\label{fig7}
\end{figure}

More precisely, we have numerically checked the above instability directly
based on Eq.(\ref{instability}) without the approximation Eq.(\ref%
{solk=omega}). It has turned out that the collapse of BEC, estimated from
the linear instability analysis, takes place just before the stagflation
point $t_{s}$. This means that a small amount of vacuum energy remains at
this point since $\rho =0$ realizes only exactly at $t=t_{s}$. However
further evolution of the uniform BEC from that point leads to the next
instability of this BEC, which yields much smaller remnant vacuum energy.
These sequence of BEC collapse continues many times and the vacuum energy
reduces at each time. Figs.(\ref{fig5}-\ref{fig7}) show a typical example of
the numerical demonstration of this sequence. Fig.(\ref{fig5}) shows the
time evolution of the BEC order parameter $\phi $. Each segment represents a
history from the generation toward the decay of BEC. The average value of $%
\phi $ reduces in time and the duration of each history becomes longer. Fig.(%
\ref{fig6}) shows the time evolution of the scale factor $a\left( t\right) $%
. Initial inflationary era, which is a straight line on this graph, is
followed by much milder expansion era. The latter is shown in the inset of
this figure and is almost the same as radiation dominated cosmic expansion.
Fig.(\ref{fig7}) shows the time evolution of BEC energy density $\rho _{\phi
}$(solid line), boson gas energy density $\rho _{g}$ (gray line), and the
Hubble parameter $H$ (dashed line). The reduction sequence is apparent. In
our demonstration Fig.(\ref{fig7}), the vacuum energy at each segment era
reduces as $1.0$, $0.0063$, $0.00072$, $0.00016$, $0.000046$, $0.000014$, $%
4.6\times 10^{-6}$, $1.5\times 10^{-6}...$ in the unit of the original
vacuum energy $V_{0}.$

The above special feature mainly comes from the stagflation, which induce
the decay of the uniform mode of BEC or classical scalar field. The
stagflation itself can be realized when the potential allows negative
region, which will be further discussed in the summary section. Thus the
stagflation and the associated reduction of the cosmological constant are
not special features in the universe.


\subsection{Generation of density fluctuations - comparison with observations%
}

As we have examined in the previous two subsections, both the reheating due
to the particle production and the decay of the vacuum energy due to the BEC
instability are most effective at around the stagflation time $t=t_{s}$.
Furthermore, we observed that this set of BEC generation and decay repeats
many times and this sequence makes the effective cosmological constant
reduce toward zero. Although the detail depends on the BEC decay dynamics
which is quite nonlinear and goes beyond the present paper, we may naturally
assume that the universe approaches toward the hot FRW universe. Then its
fate will be well described by the standard cosmology until it enters into
LA regime in most recent era. Therefore the most characteristic feature of
our model will be related to the phenomena around the stagflation point $%
t_{s}$. Then one of the remnant characteristic features to our BEC model
will be the power spectrum of the produced density fluctuations reflecting
ever increasing development of the BEC component toward the end of the
inflationary era.

We now calculate the produced density fluctuations during the inflation.
Since our BEC behaves almost like scalar field during the inflation and
therefore we can safely apply the standard analysis on the density
fluctuations. Based on the squeezed quantum state in de Sitter space with
slow-rolling approximation, we can estimate the curvature fluctuations and
the tensor fluctuations. The power index $n_{R}$ for the former is given 
\cite{Tsujikawainflation} by 
\begin{equation}
n_{R}=1-6\varepsilon +2\eta
\end{equation}%
where $\varepsilon $ and $\eta $ are defined in Eq.(\ref{epsilon and eta}).
We can calculate them based on the approximate solution $\phi _{\inf }\left(
t\right) $\ during the inflation Eq.(\ref{phi-inf})%
\begin{equation}
n_{R}=1-\frac{3\left( 1+\frac{9\mu ^{2}H_{\inf }^{2}}{16\kappa N^{2}}\right) 
}{N\left( 1-\frac{9\mu ^{2}H_{\inf }^{2}}{16\kappa N^{2}}\right) ^{2}},
\end{equation}%
which reduces, after use of Eq.(\ref{Delta-inf}), to%
\begin{equation}
n_{R}=1-\frac{3}{N}+O\left( \frac{1}{N^{3}}\right) ,
\end{equation}%
where $N$ is the e-folding number defined by Eq.(\ref{Delta-inf}) and the
next line. The tensor to scalar ratio $r=P_{T}/P_{R}=16\varepsilon $ becomes%
\begin{equation}
r=\frac{9\mu ^{2}H_{\inf }^{2}}{\kappa N^{3}\left( 1-\frac{9\mu ^{2}H_{\inf
}^{2}}{16\kappa N^{2}}\right) ^{2}}=\frac{9\mu ^{2}H_{\inf }^{2}}{\kappa
N^{3}}+O\left( \frac{1}{N^{4}}\right) .
\end{equation}%
In our model, they are estimated as 
\begin{eqnarray}
n_{R} &\approx &0.95\text{ for }N=60, \\
r &\approx &4.2\times 10^{-12}\text{ for }N=60,\mu H_{\text{inf}%
}=10^{-4},\kappa =0.1.
\end{eqnarray}%
These values are well within the observationally allowed range so far \cite%
{Komatsu09}:

\begin{eqnarray}
n_{R} &=&0.960\pm 0.013, \\
r &<&0.22.
\end{eqnarray}

We will need to examine other observational tests in the next step in order
to confirm the unified BEC cosmology. We will save the sequel of this
analysis for our future papers.

\section{Summary and Discussions}

In the previous papers, we developed the BEC cosmology to describe DE/DM in
a unified manner in LA. In this paper, we have applied this model to the
early universe and unified EA and LA without introducing new ingredients.

It turns out that the BEC cosmology well fits to describe inflationary
dynamics. Being triggered by the condensation of bosons in the environment
of huge vacuum energy, the inflation naturally initiates. This inflation
autonomously terminates due to the stagflation stage which inevitably takes
place exactly at zero energy density. At the stagflation point, particle
production and the decay of BEC occur. The former makes the universe connect
to the standard hot big bang stage and the latter guarantees the vanishingly
small cosmological constant after the inflation. Further, we have calculated
the density fluctuations produced in this model, which turns out to be in
the range allowed by the present observational data.

It should be remarked that this stagflation regime is important to solve
several problems in the early universe. Although this stagflation regime
seems to be characteristic to the BEC model, it is not true. The stagflation
is actually general and robust in various cosmological models provided that
the potential of the condensation or the inflaton can become negative. It
means that there is no necessity to adjust the potential minimum as the zero
point of energy from the beginning but it is realized autonomously through
the stagflation. The stagflation is a normal physical mechanism and does not
violate any fundamental principles. The total energy density, for example,
is guaranteed to be non-negative despite the negative potential. Furthermore
the potential unbounded from below for a uniform mode of the condensation
does not mean the catastrophe of the whole system, but only yields BEC
collapse into non-uniform modes.

\begin{figure}[h]
\begin{center}
\begin{tabular}{cc}
\resizebox{120mm}{!}{\includegraphics{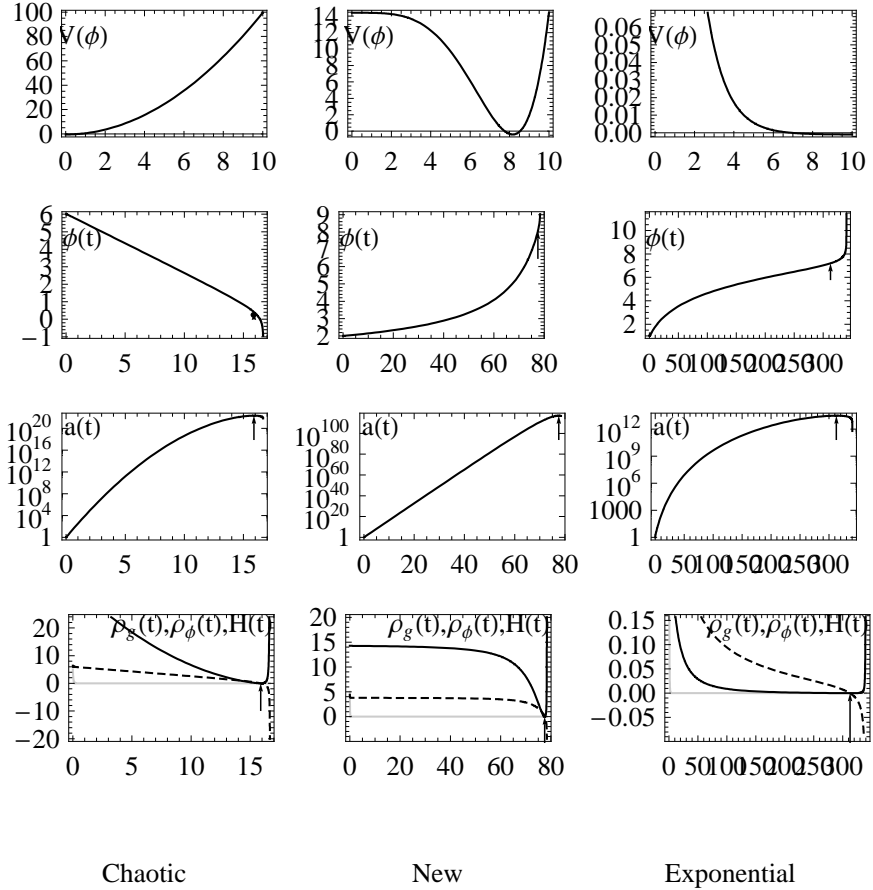}} &  \\ 
& 
\end{tabular}%
\end{center}
\caption{ Demonstration of the robustness of stagflation in various models
of inflation. The columbs show, from left to right, chaotic inflation, new
inflation and the inflation with exponential potential. The rows show, from
top to bottom, the potentials $V\left( \protect\varphi \right) $, time
evolutions of BEC order parameter $\protect\varphi \left( t\right) $, those
of scale factor $a\left( t\right) $, and those of energy densities (grey
line for $\protect\rho _{g}\left( t\right) $, solid line for $\protect\rho _{%
\protect\phi }\left( t\right) $) as well as the Hubble parameter $H\left(
t\right) $(dashed line). All of them show stagflation, market by arrows,
without any fine tuning. }
\label{fig8}
\end{figure}

In order to demonstrate that the stagflation is a general mechanism in the
universe, we pick up several typical models of inflation and slightly
generalize them by adding small negative constant term $-V_{1}$. The models
are the chaotic inflation, the new inflation, and the inflation with
exponential potential. The potentials of them are generalized to\ 
\begin{eqnarray}
V_{\text{chaotic}}\left( \phi \right) &=&\frac{1}{2}m^{2}\phi ^{2}-V_{1}, \\
V_{\text{new}}\left( \phi \right) &=&-\frac{1}{2}m^{2}\phi ^{2}+\frac{%
\lambda }{4}\phi ^{4}-V_{1}, \\
V_{\text{exponential}}\left( \phi \right) &=&e^{-\phi }-V_{1},
\end{eqnarray}%
with $m^{2}>0,\lambda >0,V_{1}>0$. We numerically calculate the fate of the
inflaton field $\phi \left( t\right) $\ in Fig.(\ref{fig8}). In this figure,
the columns show, from left to right, inflation models of chaotic, new and
exponential potential. The rows show, from top to bottom, the potentials $%
V\left( \phi \right) $, time evolutions of BEC order parameter $\phi \left(
t\right) $, time evolutions of scale factor $a\left( t\right) $, and time
evolutions of energy densities (grey line for $\rho _{g}\left( t\right) $,
solid line for $\rho _{\phi }\left( t\right) $) as well as the Hubble
parameter $H\left( t\right) $(dashed line). All of the models show
stagflation, market by arrows in the figure, without any fine tuning.

BEC cosmology seems to be able to describe the two accelerating stages of
the universe in different ways, one is the early inflationary acceleration
(EA) and the another is the contemporary mild acceleration (LA). The
dynamics of the inflation is triggered by the BEC condensation and the
instability of BEC around the stagflation point guarantees that the
cosmological constant is vanishingly small irrespective of the initial
vacuum energy. On the other hand, the contemporary mild acceleration
inevitably takes place from the balance of the condensation speed and the
potential force. Its attractor nature guarantees the existence of the tiny
cosmological constant presently observed. EA corresponds to the over-hill
regime and LA the mild-inflation regime in the general BEC cosmology\cite%
{Fukuyama06}.

These two acceleration regimes may be connected with each other by multiple
collapses of BEC, as suggested in Figs.(\ref{fig5}-\ref{fig7}). In this
paper, we have demonstrated the post-sequence of BEC collapse after
inflation and the pre-sequence of it before the final acceleration. These
two sequences may actually be connected to form a single sequence of BEC
collapse. Unfortunately, the detailed analysis on this possibility and the
mechanism of BEC collapse itself go beyond the present paper. We will figure
out the dynamics of BEC probably by numerical technique and report them in
the near future.

Novel properties of BEC cosmology in this paper is realized by the phase
transition of BEC, which cannot directly described by a unitary Lagrangian
dynamics. It will be important to notice that the universe may not be a
mechanical machine whose time evolution is unitary, but be a whole sequence
of phase transitions allowing the emergence of new structures in various
stages. This paper is also a first proposal for the resolution of the
cosmological constant problem based on the non-unitary phase transitions
including non-uniform modes of condensation. We believe the most appropriate
method to describe such incoherent evolution is the generalized effective
action formalism \cite{Morikawa95} on which we would like to report
comprehensively in our future paper including various applications and
verifications of BEC cosmology.

\section{Acknowledgements}

We sincerely appreciate fruitful discussions with Eriko Tanaka, Osamu
Iguchi, Takayoshi Ohtsuka, and Takayuki Tatekawa.

\end{document}